\documentclass[12pt]{spieman}  
\usepackage{amsmath,amsfonts,amssymb}
\usepackage{graphicx}
\usepackage{setspace}
\usepackage{tocloft}
\newcommand{\as}{\prime\prime}

\title{Pointing System for the Balloon-Borne Astronomical Payloads}

\author[a]{K.~Nirmal}
\author[a]{A.~G.~Sreejith}
\author[a]{J.~Mathew}
\author[a]{M.~Sarpotdar}
\author[a]{S.~Ambily}
\author[a]{A.~Prakash}
\author[a]{M.~Safonova}
\author[a]{J.~Murthy}
\affil[a]{Indian Institute of Astrophysics, Bangalore, India}

\cftpagenumbersoff{figure}
\cftpagenumbersoff{table} 
\begin{document} 
\maketitle

\begin{abstract}
We describe the development and implementation of a light-weight, fully autonomous 2-axis pointing and stabilization system designed for balloon-borne astronomical payloads. The system is developed using off-the-shelf components such as Arduino Uno controller, HMC 5883L magnetometer, MPU-9150 Inertial Measurement Unit (IMU) and iWave GPS receiver unit. It is a compact and rugged system which can also be used to take images/video in a moving vehicle, or in areal photography. The system performance is evaluated from the ground, as well as in  conditions simulated to imitate the actual flight by using a tethered launch. 
\end{abstract}



\begin{spacing}{2}   
\section{Introduction} 

High-altitude balloon platforms are an economical alternative to space missions for  testing instruments as well as for specific classes of observations, particularly those that require a rapid response, such as comets or other transients. We have developed a number of payloads which operate in the near ultraviolet (NUV: 200--400 nm) which we have flown on high-altitude balloons\cite{Safonova16}. We are limited to payloads under 6 kg for regulatory reasons and this constrains our payload size. Our first experiments were of atmospheric lines\cite{Sreejith16} where the pointing stability is less important,  but we do plan to observe astronomical sources for which a pointing mechanism is required. Light balloons are an exceptionally challenging platform for accurate pointing because the platform itself is in constant motion, sometimes with violent jerks and rotations. Most pointing systems for scientific balloon experiments to date have been designed for the use on large balloons with payload weights of a tonne, or more. Such systems include SPIDER\cite{Crill08}, BETTII\cite{Rizzo}, BOOMERANG\cite{Crill}, BLAST\cite{Pascale} or BLAST-Pol\cite{Fissel}. The accuracy of pointing of these systems varies from several arcminutes to few  arcseconds. For example, the pointing system in SPIDER has an accuracy of $1^{\circ}$ and in BLAST Pol of $30^{\as}$; accuracy increasing with weight and complexity of the system.

In a broad sense, pointing systems for high-altitude balloons consist of four parts: 1. attitude sensors (ASs), 2. actuators, 3. attitude control system, and 4. mechanical structure.
In this work, we describe the design and realization of a low-cost light-weight 2-axis correction pointing and stabilization system intended for use in small balloon flights, built completely using off-the-shelf components. The primary challenge in this development is that its weight must be under 1 kg, given the total mass constraint of 6 kg.

We plan to use this pointing system with other instruments that we are developing. The immediate requirements for accurate pointing come from a light-weight (650 gm) compact star-sensor camera {\it StarSense} with an accuracy of $30^{"}$ and $10^{\circ}$ field of view (FOV) \cite{Mayuresh}, and a wide-field compact ($15 \times 15\times 35$ cm) NUV imager \cite{Ambily}. Both these instruments are developed for use in small balloon payloads, as well as in nanosatellites or CubeSats, and will have a test flight in November 2016 \cite{Sreejith2016}.

\section{Basic Principle and Realization of the System}

Real-time communication with the payload on small balloons is difficult because of weight and radio-licensing constraints, and pointing mechanisms must therefore be autonomous. Our system is built with off-the-shelf electronic components and light-weight high-precision digital servomotors, where the user sets the pointing direction in the controller (Arduino Uno\footnote{\tt{http://www.Arduino.cc}}) in inertial coordinates Right Ascension (RA) and Declination (Dec) before the flight, and the pointing system is responsible for maintaining this direction regardless of balloon motion. In ground-based pointing and tracking systems, equatorial mounts are usually better suited for tracking celestial objects. However, such mounts require a fixed polar axis which is difficult to maintain in the generally unstable balloon flight. In such environment, it is easier to use an alt-az mount, where we measure the angular displacement in the vertical direction from the horizon ($0^{\circ}$ altitude or elevation), and the angular displacement in the horizontal direction  from the magnetic north ($0^{\circ}$ azimuth).

\begin{figure}[h!]
\centering
\includegraphics[scale=0.75]{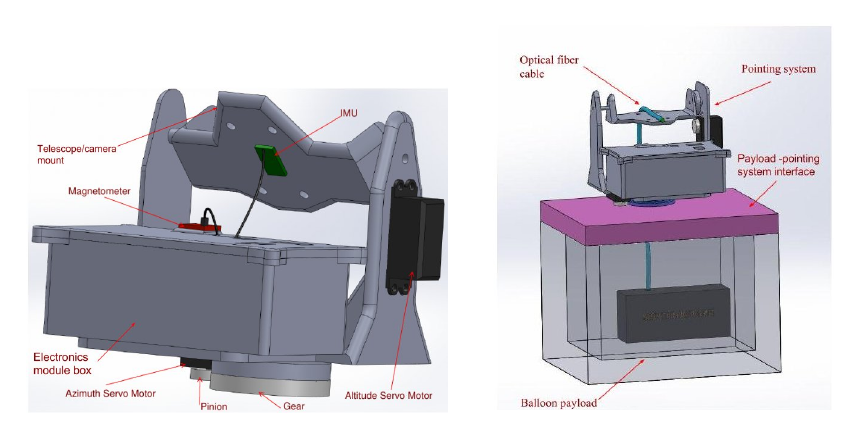}
\caption{{\it Left:} Mechanical structure of pointing system. {\it Right:} Pointing system mounted on the payload.}
\label{fig:mech}
\end{figure}

The mechanical design of the pointing system is driven by the need to keep the weight low. The structure consists of an inner frame which slews in elevation, and an outer frame which slews in azimuth (Fig.~\ref{fig:mech}, {\it Left}). Each frame is controlled by a servomotor: the inner frame uses a non-continuous servomotor and the outer uses a continuous servomotor\footnote{Savox SH-1290MG \tt{http://www.savoxusa.com}}. The shaft of these motors can be moved accurately to the desired angle using an internal electronic circuit, which identifies the current angle of the motor shaft from a reference point and then moves the shaft to the desired position. The rotation is limited to $0^{\circ}-85^{\circ}$ range to avoid `gimbal locking' of the IMU\cite{Savvidis} used to determine the attitude (Fig.~\ref{fig:mech}, {\it Left}). However, the outer frame requires rapid rotation from $0^{\circ}$ to $360^{\circ}$ in azimuth, thus requiring a continuous servomotor\footnote{Dynamixel MX-28T \tt{http://www.trossenrobotics.com}}.

The mechanical and structural design of the system (inner and outer frames, Fig.~\ref{fig:mech}, {\it Left}) was performed with SolidWorks 3D design software\footnote{\tt{http://www.solidworks.in}}, and the structure was printed on a Replicator2 Desktop 3D printer from MakerBot\footnote{\tt{http://www.makerbot.com}}
using polylactic acid filament (PLA) --- a biodegradable polymer produced from lactic acid. PLA is harder and melts at a lower temperature ($180^{\circ}$ to $220^{\circ}$C) than ABS (Acrylonitrile Butadiene Styrene), another popular material used in 3D printers which has a glass transition temperature of $105^{\circ}$C. The complete balloon payload--pointing system assembly is shown in Fig.~\ref{fig:mech}, {\it Right}.

\section{Control Mechanisms}

\begin{figure}[h]
\begin{center}
\includegraphics[scale=0.5]{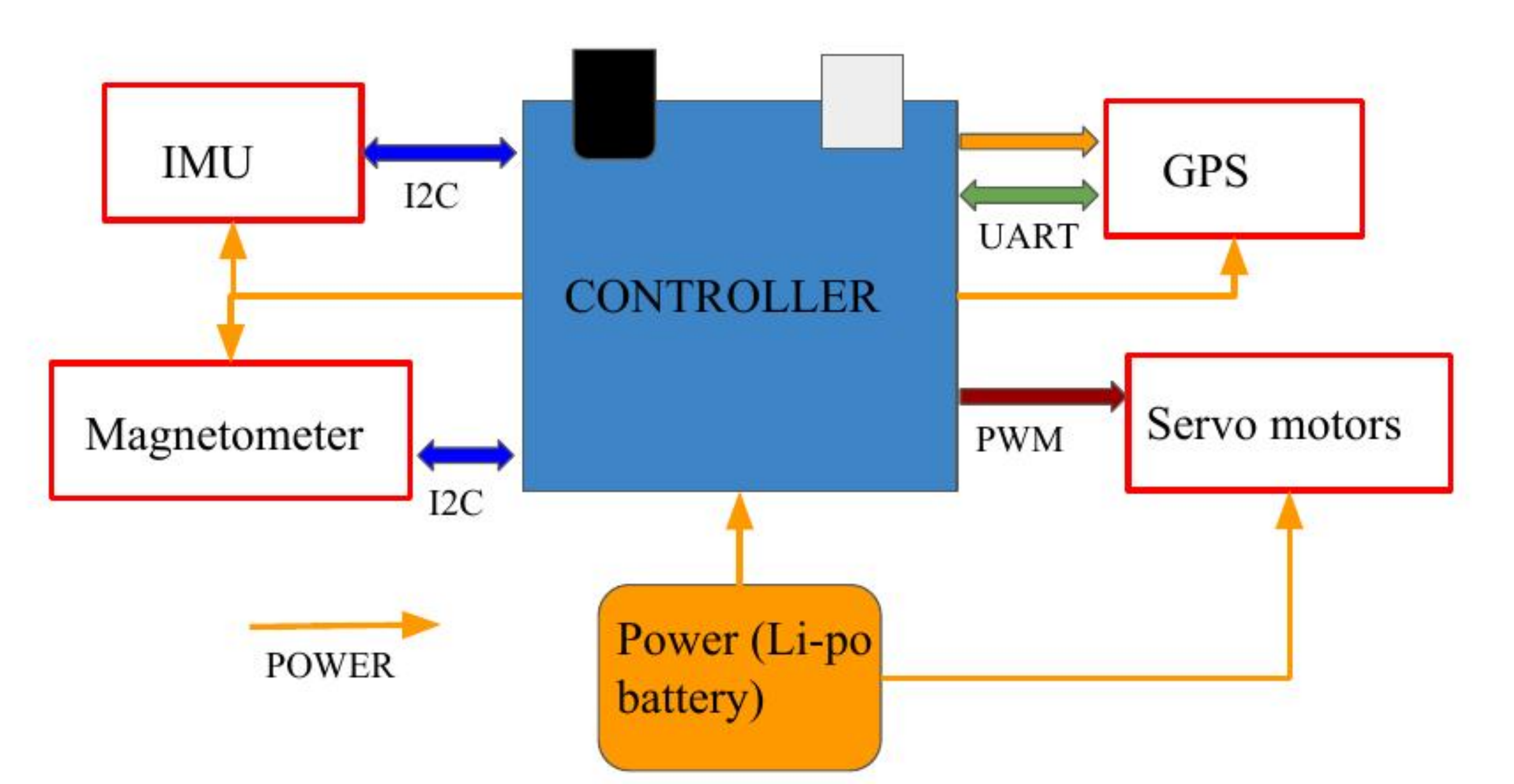} 
\end{center}
\caption{A block diagram of the Attitude Control System (ACS).}
\label{fig:acs}
\end{figure}

The Attitude Control System (ACS) (Fig.~\ref{fig:acs}) comprises an Arduino Uno controller, a set of microelectromechanical system (MEMS) sensors, actuators and a GPS unit\footnote{iWave Systems SiRF StarIII GSC3f GPS receiver, iWave Systems, India (\tt{http://www.iwavesystems.com})}. The Arduino Uno controller is an open-source electronic platform based on easy-to-use hardware and software, developed at the Interaction Design Institute Ivrea, Italy. We have chosen this controller over its alternatives (Teensy\footnote{{\tt https://www.pjrc.com/teensy}.}, BeagleBone\footnote{{\tt http://beagleboard.org/bone}.} and Raspberry Pi\footnote{{\tt https://www.raspberrypi.org}.}) because of its extensive library of available software\cite{Ham}.

The function of the ACS can be divided into three parts: 
\begin{enumerate}
\item Finding the pointing direction;
\item Estimation of the current position of the pointing system in terms of the azimuth and the elevation using IMU and magnetometer output;
\item Calculating the difference between the desired and the actual pointing directions, and moving the platform (telescope) to the desired position.
\end{enumerate}
The first step is performed by converting the user-provided equatorial coordinates  (RA and Dec), into elevation (ALT) and azimuth (AZI) using Eq.~\ref{eq:alt_az}.

\subsection{ACS computation}

The accuracy of ACS computation was checked  by calculating the position of the Sun every second for about 30 minutes, and 
comparing it with the actual values. The equatorial coordinates of the Sun do not change noticeably over the duration of the 
observation, but the position in the sky will change by $8^{\circ}$ due to the Earth's rotation, and its apparent position will 
change by another $0.833^{\circ}$ due to atmospheric refraction.  We programmed the initial position of the Sun, time, date and 
location of the observation (Table~\ref{table:sunPosition}) into the ACS. The values of altitude and azimuth calculated by the 
controller using Eq.~\ref{eq:alt_az} were compared with the actual values obtained from NOAA\footnote{Data provided by NOAA ESRL 
Global monitoring Division, Boulder, Colorado, USA (\tt{http://esrl.noaa.gov/gmd}).}. The errors in this 
calculation were $\pm 0.006^{\circ}$, within our desired precision. 

\begin{table}[h]
\centering
\caption{ACS programmed parameters.}
\begin{tabular}{ll} \hline
Date &  18/06/2015\\ 
Time (IST) &  8:00 am\\ 
Location &  Hoskote \\ 
Latitude &   $13.113^{\circ}$ N\\ 
Longitude &  $77.811^{\circ}$ E\\ 
Programmed RA of Sun &  $86.269^{\circ}$ \\ 
Programmed Dec of Sun &  $23.390^{\circ}$\\ 
\hline
\end{tabular}
\label{table:sunPosition}
\end{table}

\subsection{Estimation of the attitude}

We placed an IMU (MPU-9150) on the inner frame (Fig.~\ref{fig:mech}, {\it Left}) to measure the elevation of the pointing system. The MPU-9150 comprises an in-built 3-axis magnetometer (AK8975), a combination of a 3-axis accelerometer and a 3-axis gyroscope (MPU-6050), and a digital motion processor capable of motion fusion\footnote{IMU gives the Euler angles (roll, pitch and yaw) of rotation as an output.}. A better accuracy is achieved by fusing the data from the individual sensors\cite{Emilson}, and we have used accelerometer and gyroscope combined output to measure the elevation.
Although this IMU includes a magnetometer, it moves in elevation due to its location on the inner frame and thus the its readings are unreliable. We, therefore, mounted another magnetometer (HMC5883L) on the outer frame (Fig.~\ref{fig:mech}, {\it Left}), which moves around an axis perpendicular to the Earth's magnetic field, to measure the azimuth.

This magnetometer consists of high-resolution magneto-resistive sensors with an application-specific integrated circuit, containing an amplifier, automatic degaussing strap drivers, and offset cancellation circuits. The analog data from the magneto-resistive elements is digitized using an in-built 12-bit ADC. Any drift in the sensor measurements can be calibrated out by using its self-test mode, which internally excites the sensor with a nominal magnetic field.

The magnetometer and the IMU are connected to the Arduino I2C port, which is a multi-master serial single-ended computer bus through which low-speed peripherals are attached to the controller. The in-built MEMS gyroscope consists of vibrating solid state resonators that maintain their plane of vibration even if the gyroscope is tilted or rotated. This type of gyroscope is known as Coriolis vibratory gyroscope (CVG), and is common in consumer electronics such as tablets and mobile phones. A voltage, proportional to the angular velocity of the IMU, is generated in the gyroscope and is digitized using a 16-bit ADC, whose full scale reading corresponds to 3.3 V. Thus, the analog voltage generated for an angular velocity of $1^{\circ}$/sec is 3.3  mV/($^{\circ}$/sec)\cite{mpu_datasheet}, corresponding to an ADC value of  $\frac{3.3\,mV}{3.3\,V} \times (2^{16}-1 ) = 65.535$. The gyroscope generates a bias voltage, which is measured when the IMU is stationary, and this bias is subtracted from the ADC value to get the actual response. The angular velocity rate in degrees per second is calculated using the following formula:
\begin{eqnarray}
\mbox{Angular velocity rate} (\omega) &=& \frac{(V_{\rm ADC} - V_{\rm bias})}
{\text{sensitivity}}\,  (^{\circ}/sec) \,,
\label{eq:omega}
\end{eqnarray}
and the angular displacement is calculated  by multiplying the rate by the time period $\Delta t$ ($\theta = \omega \times \Delta t$). 

The accelerometer measures the acceleration (g) in $X$, $Y$ and $Z$ axes and generates a voltage proportional to the acceleration, which is digitized by the 16-bit ADC and read by the Arduino controller through its I2C port. The bias voltage ($1.5\,V$), which is inherent in the accelerometer ADC output, is subtracted from the accelerometer output to get the voltage corresponding to the acceleration. The elevation in degrees is found from acceleration values using the equation
\begin{eqnarray}
ALT = \text{arctan}\left(\frac{{V_{accy}}}
{{V_{accz}}}\right) + \pi\,,
\end{eqnarray}
where $V_{accy}, V_{accz}$ are the bias-subtracted accelerometer ADC outputs corresponding to acceleration in $Y$ and $Z$ axes, respectively.

\begin{figure}[ht]
\centering
\includegraphics[scale=0.3]{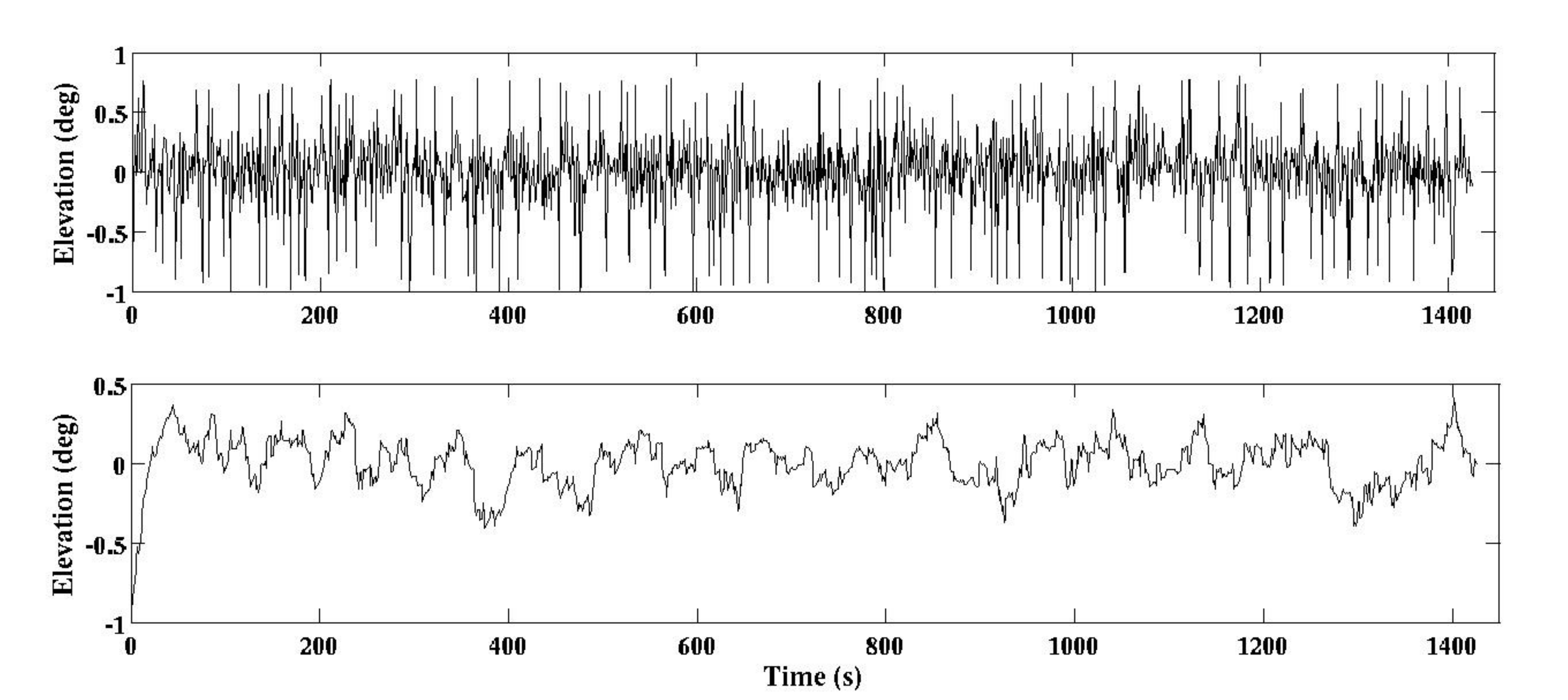}
\caption{\textit{Top}: Elevation calculated from the accelerometer shows considerable scatter. \textit{Bottom:} The sensor-fused elevation data from accelerometer and gyroscope using Kalman-filter is much smoother.}
\label{fig:sensfusion}
\end{figure}

The gyroscope gives precise values over short time duration but drifts for longer observations\cite{Sreejith14}, while the converse is true for the accelerometer: there drift is negligible over long periods of time but a significant jitter  occurs on short time scales ($0.330^{\circ}$ in 100 ms bins). We can reduce this jitter by combining the data from the gyroscope and the accelerometer using a Kalman filter\footnote{{\tt https://github.com/TKJElectronics/}.}. This is shown in Fig.~\ref{fig:sensfusion}, where the elevation, calculated from only the accelerometer, is plotted in the top panel, and the elevation from the fusion of the two sensors is plotted in the bottom panel. The jitter in the elevation is reduced to $0.143^{\circ}$ per 100 ms.

We obtained the azimuth from the external magnetometer (HMC5883L). The magnetic field is read using the intrinsic HMC5883 library, which gives the Earth's magnetic field vectors in $\mu$T in $X$, $Y$ and $Z$ axes (Earth-centered inertial frame). We then calculated the azimuth using the following equation,
\begin{eqnarray}
AZI = \text{arctan} \left(\frac{-\text{magnetic vector field strength in $Y$ axis}}
{\text{magnetic vector field strength in $X$ axis}}\right)\,,
\end{eqnarray}
which was implemented using the Adafruit Unified Sensor Library\footnote{{\tt http://www.adafruit.com/}.}. We corrected the azimuth for the declination angle --- the difference between the magnetic North and the true North for our location (Hoskote latitude $13.11277^{\circ}$ N, Longitude $77.8113^{\circ}$ E and $-1^{\circ} 33'$ declination angle) --- in the code.

\section{Correcting the pointing position}

Prior to  the implementation of a control system for pointing position correction, an estimate of the amount of the payload motion during the flight is required. Because light balloons are in constant motion, the payloads are usually subject to violent jerks and rotations, and other disturbances that the pointing control must reject. We have flown our attitude sensor (AS) in different balloon flights, and estimated the payload motion in roll, pitch and yaw axes (tilt, elevation and azimuth) during the flights. We inferred the prerequisite of the pointing correction mechanism from this data, and simulated a realistic model of our system in MATLAB. We have implemented a proportional-integral-derivative (PID) loop feedback algorithm in the simulation, and in the real system, to correct for the difference between the actual pointing and the desired pointing in a closed loop. We found that the results obtained from the simulation and the actual measurements were comparable.

\subsection{Estimation of disturbances on the payload}

\begin{figure}[h]
\centering
\includegraphics[scale=0.6]{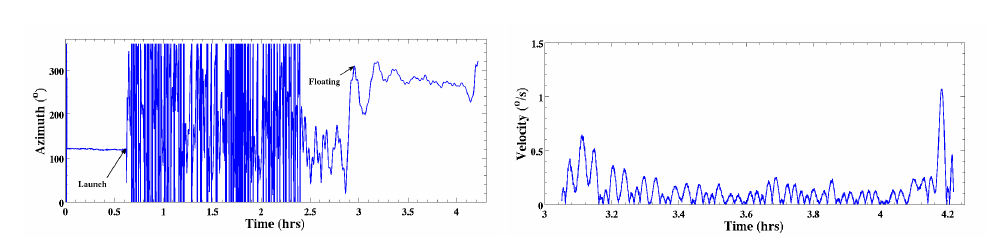}
\caption{{\it Left:} The motion of the payload in azimuth during the balloon flight. {\it Right:} The rms velocity ($^{\circ}/s$) of the payload in azimuth.}
\label{fig:disturb}
\end{figure}

We inferred from the AS data that our balloon flights  have a turbulent phase while ascending, and a stable phase at high altitudes (above 19 km)  (Fig.~\ref{fig:disturb}), especially at float altitudes ($\sim 30$ km)\cite{ritareport}. The motion of the payload during these stages is tabulated in Table.~\ref{tbl:tifr}. The prerequisite parameters for our control system, such as settling time, maximum peak overshoot and steady state error, are derived from this data. 

\begin{table}[h]
\centering
\caption{Motion of the payload in elevation, azimuth and tilt.}
\vspace{0.1in}
\begin{tabular}{lp{20 mm}p{20mm}} \hline
Axis & Condition &RMS velocity ($^{\circ}/s$) \\ 
\hline
Azimuth & Turbulent \newline Floating  &  25.74 \newline 0.20  \\
Elevation &  Turbulent \newline Floating &   1.35 \newline 0.01 \\
Tilt & Turbulent \newline Floating &   3.87 \newline 0.01  \\
\hline
\end{tabular}
\label{tbl:tifr}
\end{table}

\begin{enumerate}
\item The settling time of the system is 2 s.
\item The steady state error of the system is zero.
\item The maximum peak overshoot is less than 20$\%$ of the required value.
\end{enumerate}

\subsection{Simulation of control system}

\begin{figure}[h!]
\centering
\includegraphics[scale=0.5]{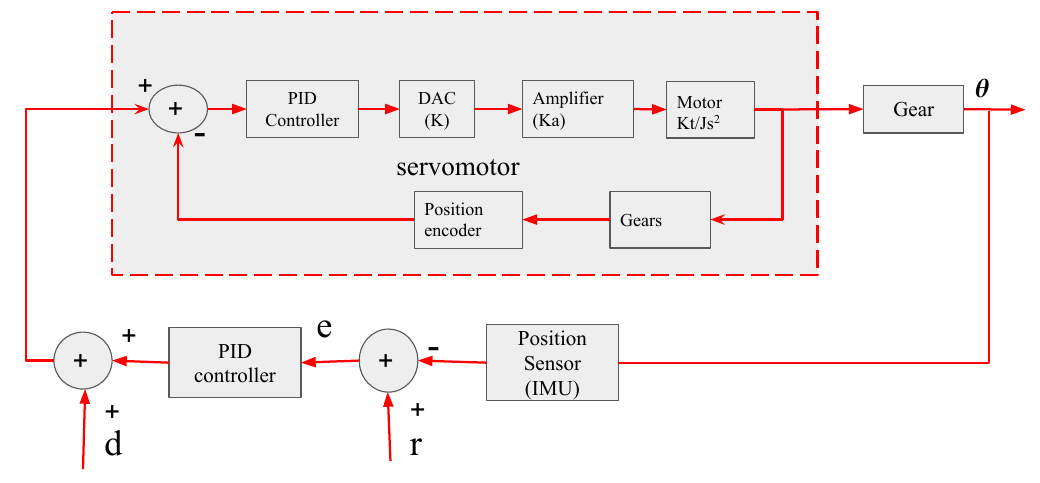}
\caption{A block diagram of control system model, where {\bf r} is the control signal, {\bf d} is the disturbance (load), {\bf e} is the error signal and $\theta$ is the angle rotated by the motor shaft.}
\label{fig:control_block}
\end{figure}

\begin{figure}[h!]
\centering
\includegraphics[scale=0.8]{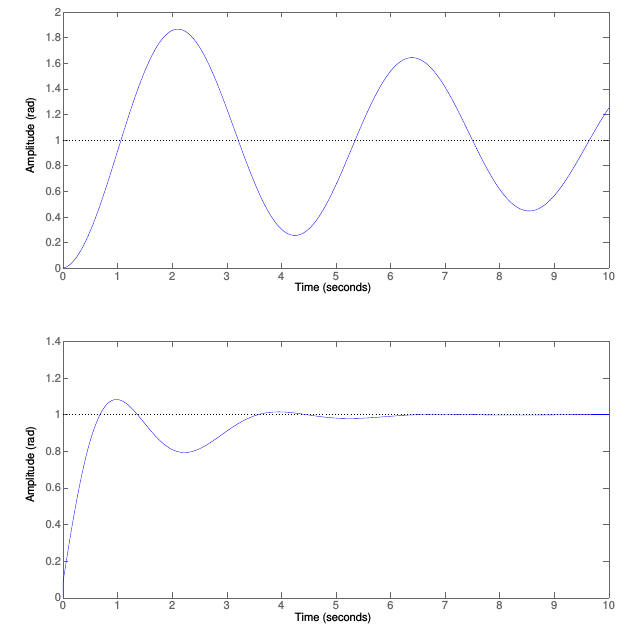}
\caption{The closed loop step response of the system. {\it Top}: without the PID controller, {\it Bottom}: with the PID controller of gain $K_p =2$, $K_d = 1$ and $K_i = 2$. The settling time is reduced to the satisfactory value.}
\label{fig:clp_pid_laod}
\end{figure}

Figure~\ref{fig:control_block} shows control system model of our pointing system. The model constitutes an actuator and external gears, and the load acting on the system. We used servomotor as an actuator in our system. Hence, a realistic model of a  servomotor was simulated (Eq.~\ref{eq:servo_tf}), considering different internal parameters such as encoder gain, gear ratio, and DAC gain. The parameters of the servomotor used in Eq.~\ref{eq:servo_tf} are tabulated in Table~\ref{tbl:servo_para}. The transfer function of the system is obtained after substituting all other external parameters, including gears and load (Eq.~\ref{eq:tf_load}). The closed-loop step response (the response of the system when the motor shaft is forced to move by 1 rad) of the system is shown in Fig.~\ref{fig:clp_pid_laod}, {\it Top}. However, this response did not satisfy our design requirements (Table~\ref{tbl:clp_load}), therefore, an external proportional-integral-derivative (PID) algorithm was introduced in the feedback path of the system (Fig.~\ref{fig:control_block}) to correct for the difference between the actual pointing and the desired pointing in a closed loop. 

\begin{equation}
\frac{\theta(s)}{V(s)} = \frac{K_t \times G \times K_{G} \times K_{DAC}}{s\left(\left(J_r s+b\right)\left(Ls+R\right)+K_t^2\right)}\,.
\label{eq:servo_tf}
\end{equation}

\begin{table}[h]
\centering
\caption{ACS programmed parameters.}
\vspace{0.1in}
\begin{tabular}{lll} \hline
$J_r $& Moment of inertia of the rotor &  $0.993 \times 10^{-7}$ kg\,m$^2$  \\
$b$ & Motor viscous friction constant &  $ 72.4 \times 10^{-6}$ N\,m\,sec\\    
$K_b$ & Electromotive force constant &  0.011 V/rad/sec \\
$K_t$ & Motor torque constant & 0.0112 N\,m/A \\           
$R$ & Electric resistance &  11.4 $\Omega$ \\               
$L$ & Electric inductance &  $343 \times 10^{-6}$ H\\
$J_l$ & Moment of inertia of the load & $903.01 \time 10^{-6}$ kgm$^2$  \\
$J$ & $J_r + J_l$ &  $903.1 \time 10^{-6}$ kgm$^2$ \\
$G_{in}$   & Internal gear ratio &  1/193 \\
$G_{ex}$   & External gear ratio &  1/2 \\
$G$   & Total gear ratio &  1/386 \\
$K_G$ & Encoder gain &  651.8 unit/rad \\
$K_{DAC}$ & Gain of DAC &  0.0468 V/unit \\ 
\hline
\end{tabular}
\label{tbl:servo_para}
\end{table}

The system response depends on the values of the constants $K_p$, $K_i$ and $K_d$ -- gains of the proportional, integral and differential error, respectively. In our simulation, we have tested   different gain values, and obtained a satisfactory result (Table~\ref{tbl:clp_load}) for $K_p =2$, $K_d = 1$ and $K_i = 2$ for the load simulating the real load intended on the system (star sensor of $\sim 500$ g). The comparison of the step response of the system without the external PID controller and with the controller is given in Fig.~\ref{fig:clp_pid_laod}. 

\begin{table}
\centering
\caption{The closed loop step response parameters without and with an external PID}\label{tbl:clp_load}
\vspace{0.1in}
\begin{tabular}{lcc} 
\hline
Parameter           & no PID & with PID \\
\hline
Rise Time (sec)     & 0.73  &  0.53 \\
Settling Time (sec) & 56    &  5.4 \\
Overshoot (\%)      & 86.3    &  8.2 \\           
Peak (rad)          &  1.86  &  1.1  \\
\hline
\end{tabular}

\end{table}

\begin{equation}
\frac{\theta(s)}{V(s)} = \frac{887 \times 10^{-6}}{3.098\times10^{-07} s^3 + 0.0103 s^2 + 0.0009509 s}.
\label{eq:tf_load}
\end{equation}

\subsection{Hardware implementation of the PID algorithm.}

\begin{figure}[h!]
\begin{center}
\includegraphics[width=5in]{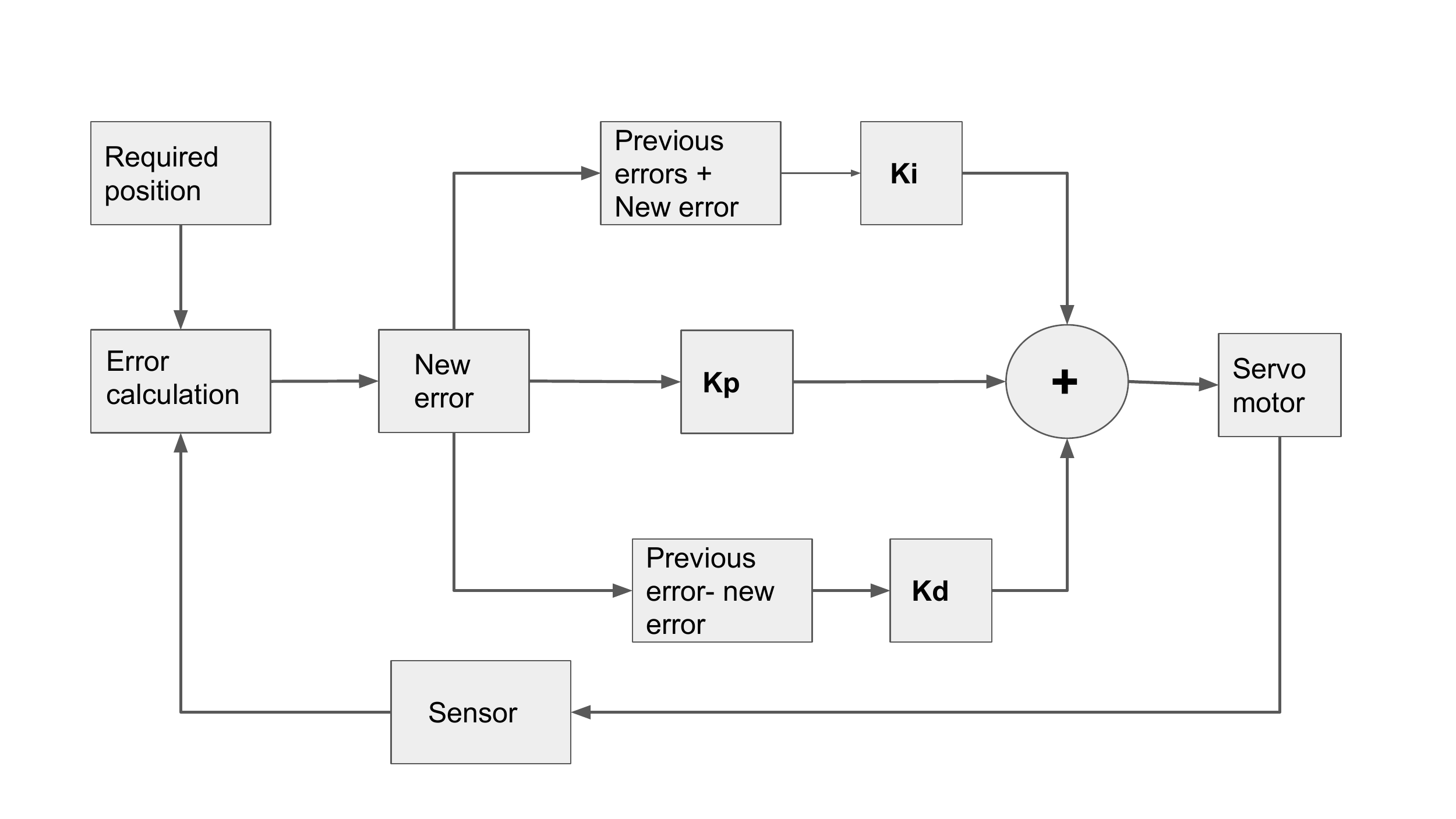} 
\end{center}
\caption{Block diagram of the PID algorithm implemented in the controller.}
\label{fig:pid}
\end{figure}

\begin{figure}[h!]
\begin{center}
\includegraphics[width=0.9\textwidth]{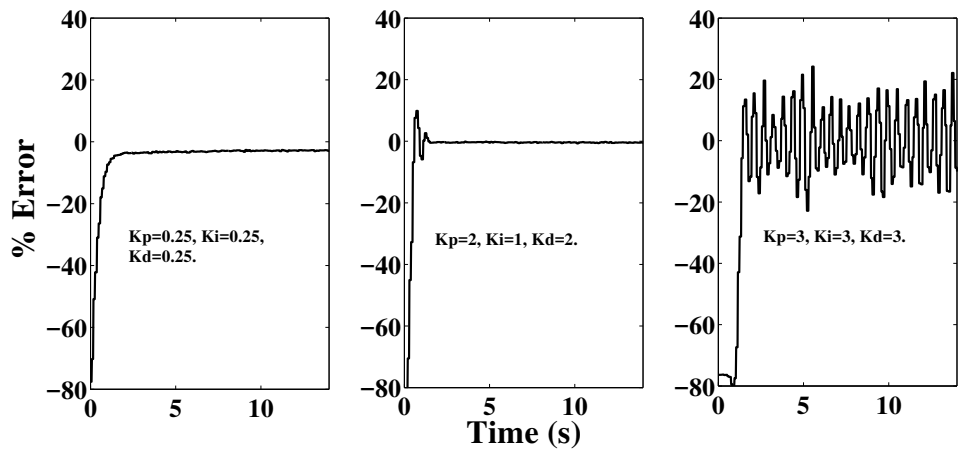}
\end{center}
\caption{The response of the PID controller for different proportional, integral and differential gains. $Y$-axis shows the deviation in percentage between the programmed and current position. For $K_p=0.25$, $K_d=0.25$, $K_i=0.25$, the system response is slow ({\it Left}). Fast and stable response achieved with gain values $K_p=2$, $K_i=1$, $K_d =2$ ({\it Middle}). For gains $K_p=3$, $K_d=3$, $K_i=3$, the system becomes unstable and starts oscillating ({\it Right}).}
\label{fig:pidK}
\end{figure}

We implemented a PID control algorithm in Arduino controller as depicted in Fig.~\ref{fig:pid}. The step response of the system for different PID gains is given in Fig.~\ref{fig:pidK}. Fast and stable response achieved with gain values $K_p=2$, $K_i=1$, $K_d =2$ ({\it Middle}), similar to the gain values obtained from the modelling under load condition.

\section{Performance tests}

The calibration and testing of the individual sensors and the attitude sensor as a complete unit has been discussed in \cite{Sreejith14}. In the next sections, we describe the testing of the pointing system for pointing accuracy and stability performance, both on the ground and in tethered flights. Pointing accuracy is determined as how accurately the system points towards the actual position of an object. Pointing stability (precision) is determined as how well the system maintains the pointing position over time.
 
\subsection{Ground test}

Performance test of our pointing system was done under controlled conditions on the ground. We mounted a camera and an external IMU (x-IMU\footnote{http://www.x-io.co.uk/products/x-imu/.}) on the inner frame of pointing system and programmed the system to point towards different celestial objects (Sun, Moon, Sirius and Jupiter) and tracked them for a duration of at least 15 minutes, with the camera recording a video (30 frames per second) of the object under observation. We performed this test on, at least, three nights for each object. Pointing accuracy of the system was found by comparing the pointing direction (elevation and azimuth ) calculated by controller and executed by the pointing system, and the direction measured by the x-IMU. The RMS pointing accuracy was within $\pm 0.28^{\circ}$. 

Stability of the system (pointing precision) was determined by taking the centroid of an object's image (from isophotes) in each 1.5 sec frame, and finding the shifts (in degrees) between the centroid in the first frame (reference frame) and in all subsequent frames. The average centroid shift  for all objects was found to be $\pm 0.13^{\circ}$. 

\begin{figure}[h!]
\centering
\includegraphics[scale=0.3]{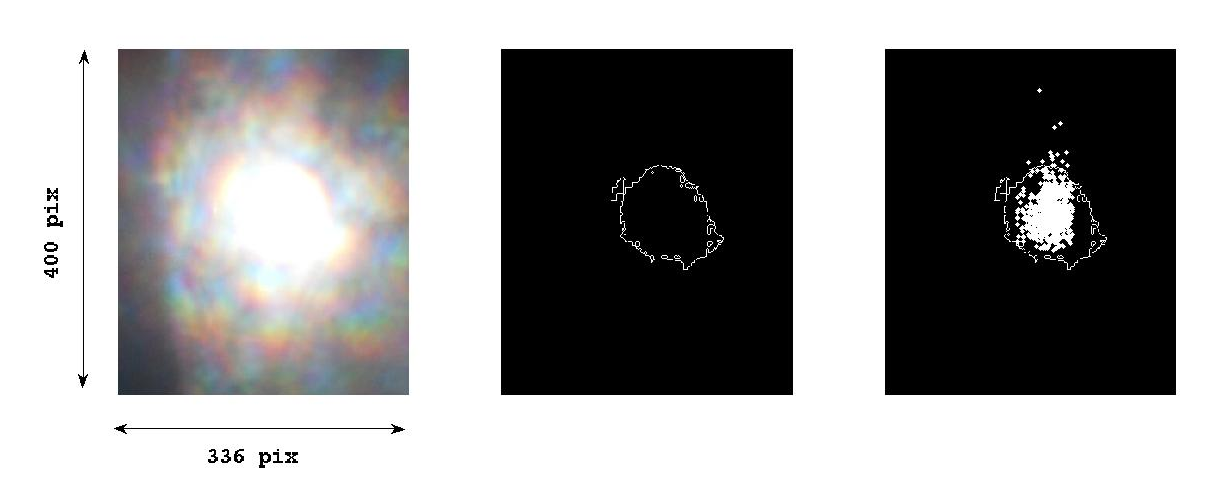}
 \caption{\textit{Left:} An image (the Sun) frame extracted from the video capture during the ground test. \textit{Middle:} The isophote of the Sun's image scaled by 250:1. This isophote was used to find the centroid of the image. {\it Right}: The isophote of the Sun with overplotted centroids calculated for every image frame. The size of an image is marked in pixels.}
\label{fig:sun}
\end{figure}

In Fig.~\ref{fig:sun}, we show an example image of the Sun ({\it Left}), the isophote ({\it Middle}) of the image, and the shifts between the centroids ({\it Right}). The RMS of the centroid shift for every tracked object is given in the last column of Table~\ref{table:groundtestresults}. 

\begin{table}[h!]
\caption{Ground test results of tracking stability}
\vskip 0.1in
\centering
\begin{tabular}{lllll} 
\hline
 & Object& Actual RA & Actual Dec & \begin{tabular}{@{}c@{}} Centroid \\ shift $ (\pm)$ \end{tabular}\\
 \hline  \\
\begin{tabular}{@{}c@{}}Test1 \\ $23/03/2016 $\end{tabular}& \begin{tabular}{@{}c@{}c@{}c@{}}Moon \\ Sirius \\ Sun \\ Jupiter\end{tabular} & 
\begin{tabular}{@{}c@{}c@{}c@{}} $ 185.223^{\circ}$ \\ $ 101.284^{\circ}$\\ $ 2.531^{\circ}$ \\ $167.193^{\circ}$ \end{tabular}& 
\begin{tabular}{@{}c@{}c@{}c@{}} $-0.853^{\circ}$ \\ $ -16.722^{\circ}$ \\ $ 1.099^{\circ}$ \\ $ 7.170^{\circ}$ \end{tabular} &
\begin{tabular}{@{}c@{}c@{}c@{}} $0.129 ^{\circ}$ \\ $  0.172^{\circ}$ \\ $ 0.119^{\circ}$ \\  $ 0.112^{\circ}$\end{tabular} \\
\hline\\
\begin{tabular}{@{}c@{}}Test2 \\ $24/03/2016 $\end{tabular}& \begin{tabular}{@{}c@{}c@{}c@{}}Moon \\ Sirius \\ Sun \\ Jupiter\end{tabular} & 
\begin{tabular}{@{}c@{}c@{}c@{}} $ 196.601^{\circ}$ \\ $ 101.284^{\circ}$\\ $ 3.443^{\circ}$ \\ $167.193^{\circ}$ \end{tabular}& 
\begin{tabular}{@{}c@{}c@{}c@{}} $-4.633^{\circ}$ \\ $ -16.722^{\circ}$ \\ $ 1.490^{\circ}$ \\ $ 7.170^{\circ}$ \end{tabular} &
\begin{tabular}{@{}c@{}c@{}c@{}} $ 0.153^{\circ}$ \\ $  0.101^{\circ}$ \\ $ 0.168^{\circ}$ \\  $ 0.100^{\circ}$\end{tabular} \\
\hline
\end{tabular}
\label{table:groundtestresults}
\end{table}

\subsection{Tethered flights}

We tested pointing accuracy and stability of our pointing system in three tethered flights (Appendix C). We pointed towards either the Sun or the Moon, depending on the time of the day. While these launches do simulate actual flights, they are buffeted by stronger winds than at the high altitudes we normally fly\cite{Manchanda}, and thus put a  greater stress on the pointing system. In each flight, the payload included the pointing system mounted on a star-shaped platform, a camera mounted on the pointing system, and an IMU (x-IMU), mounted on the platform. The camera was programmed to take an image every 5 sec, and the data were stored on-board along with the IMU data (Euler angles of the payload position during the flight). 

We have classified these tethered flights into two phases: stable and unstable, based on the motion of the platform. If the payload motion was below $25^{\circ}$/sec in azimuth and $4^{\circ}$/sec in elevation, we classified it as stable, and as unstable otherwise. The winds were particularly bad in one of the flights and several times the payload hit the surrounding trees and even the ground. However, the pointing system continued to perform throughout the total duration of the test (though sometimes the rotational motion of payload was reaching $90^{\circ}$/sec due to  strong winds).

\begin{figure}[h]
\centering
\includegraphics[scale=0.35]{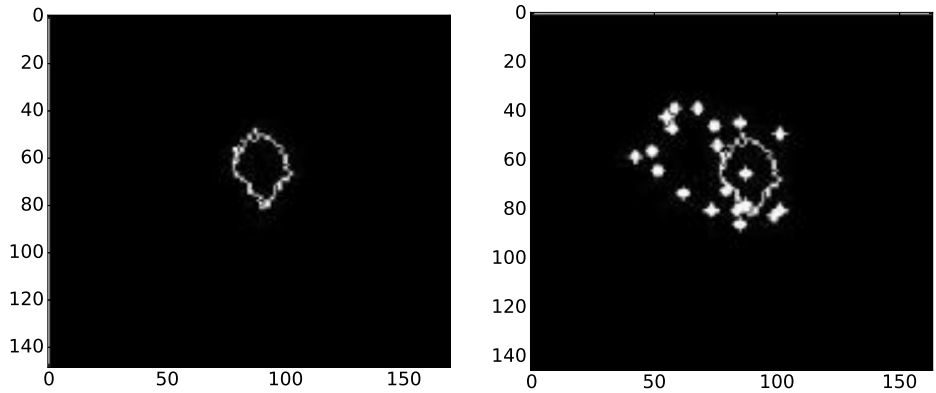}
\caption{{\it Left}: Isophote of the Sun's image obtained during a tethered flight. {\it Right}: Individual centroids of the Sun in each of the 5 second frames are plotted on the  over a period of 150 seconds. Scaling is the same as in Fig.~\ref{fig:sun}.}
\label{fig:teth_cent}
\end{figure}

The images taken during above mentioned periods were identified from the image time stamp. In order to find the stability of pointing, we applied the same method as in the ground test: took the first frame as the reference frame, and found the Sun's image centroid. The stability was measured by the shifts (in degrees) in the image centroid on every successive frame with respect to the reference frame (Fig.~\ref{fig:teth_cent}). In the stable phase, the RMS of the centroid shifts was $0.844^{\circ}$. This was largely due to the motion of the payload in the third axis (here, tilt), which contributed disproportionately to the pointing precision. During the periods when this motion was small, less than $0.1^{\circ}$ (the least winds), the stability improved to within $0.4^{\circ}$.
This pointing system is the first step towards building a star sensor-based pointing system, where the system described in the paper acts as a coarse correction for pointing direction, and fine correction will be done after the input from the star sensor which has wide FoV of 10 degrees.  Hence, the function of this system is initially to get the object inside the FoV of the star sensor camera. Therefore we find the performance of the system satisfactory.

\section{Conclusion}

We have designed and developed a low-cost light-weight, closed-loop pointing and stabilization platform for use in balloon-borne astronomical payloads. This system was build completely from off-the-shelf components: an MPU-9150 IMU, a HMC5883L magnetometer, an Arduino controller and a SiRF StarIII GSC3f GPS receiver unit. The system performance was checked on the ground and in tethered flights with satisfactory results. The system can point to an accuracy of $\pm 0.28^{\circ}$ and track objects from the ground with an accuracy of $\pm 0.13^{\circ}$. The performance in the tethered flights was poorer ($0.40^{\circ}$ in best conditions), largely because of strong winds at low altitudes. However, the stability of pointing was still within $\sim 1.6^{\circ}$ even in worst conditions. Such winds are not present in the stratosphere\cite{Manchanda}, where payloads are known to be stable at float\cite{ritareport}, and we expect pointing accuracy and stability of our system to be similar to those on the ground.

We are exploring several avenues to further improve the system performance including using better sensors and servomotors with finer steps. We have developed a star-senso\cite{Mayuresh}r with a resolution of $30^{\prime\prime}$ which we will patch into the pointing system. We plan to have a high-altitude floating balloon flight in November 2016 with an imager and a spectrograph where this pointing system will be put to use\cite{{Ambily},{Sreejith2016}}.

\section*{Acknowledgments}
We thank Prof.~D.~K.~Ojha and the staff of the TIFR Balloon Facility, Hyderabad, for allowing us to use our IMU in their flight and for sharing the GPS and flight information.

\bibliography{ref} 
\bibliographystyle{spiejour}

\appendix
\section{Pointing System Program Flowchart}

\begin{figure}[h]
\begin{center}
\includegraphics[scale=0.6]{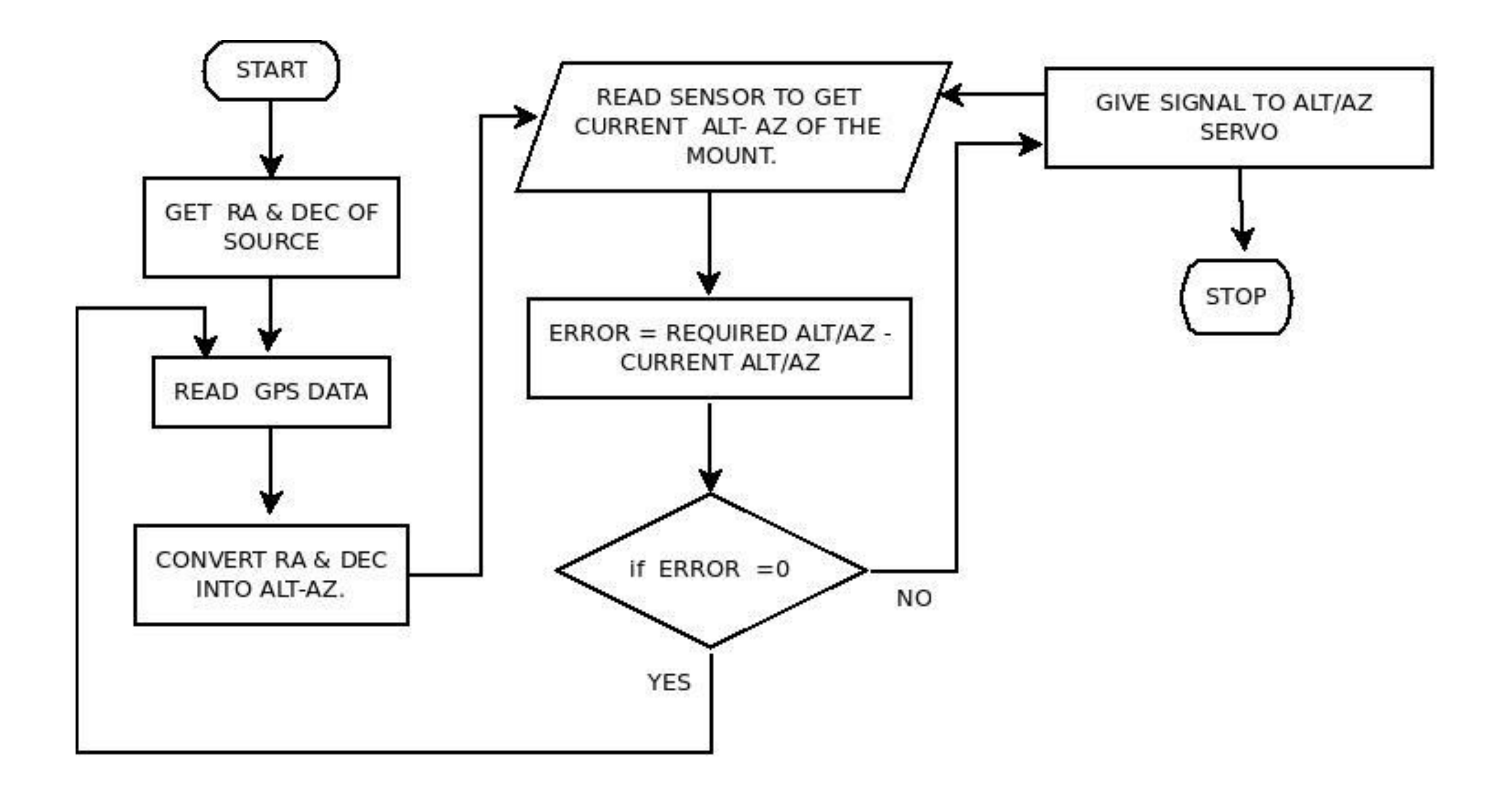} 
\end{center}
\caption{Program flowchart.}
\label{fig:flowchart}
\end{figure}

The flowchart (Fig.~\ref{fig:flowchart}) shows the program flow inside the controller. All the programming is done in Arduino platform. 

\section{Conversion of Equatorial coordinates into horizontal coordinates}

In our controller we converted the RA and Dec of the source into elevation and azimuth using the following formula,
\begin{eqnarray}
\sin{(ALT)} &=&  \sin{(Dec)}  \sin{(LAT)} + \cos{(Dec)}  \cos{(LAT)} \cos{(HA)}\,, \nonumber\\ 
\cos{(AZI)} &=& \dfrac{\sin{(Dec)} -
\sin{(ALT)}  \sin{(LAT)}}{\cos{(ALT)}  \cos{(LAT)}}\,, 
\label{eq:alt_az}
\end{eqnarray}
where LAT and HA are the latitude and the hour angle\footnote{Hour Angle -- time elapsed after a celestial body transited over observer's meridian. It is expressed in terms of local sidereal time (LST) and RA as $HA = LST - RA$.}, respectively.  The controller updates the calculation of the desired azimuth and elevation every second, using the latitude and longitude of the platform as determined by the on-board GPS.

\section{Description of Tethered Launch Experiment}

We simulated conditions similar to the balloon flight using a tethered launch.

\begin{figure}[h]
\centering
\includegraphics[scale=0.20]{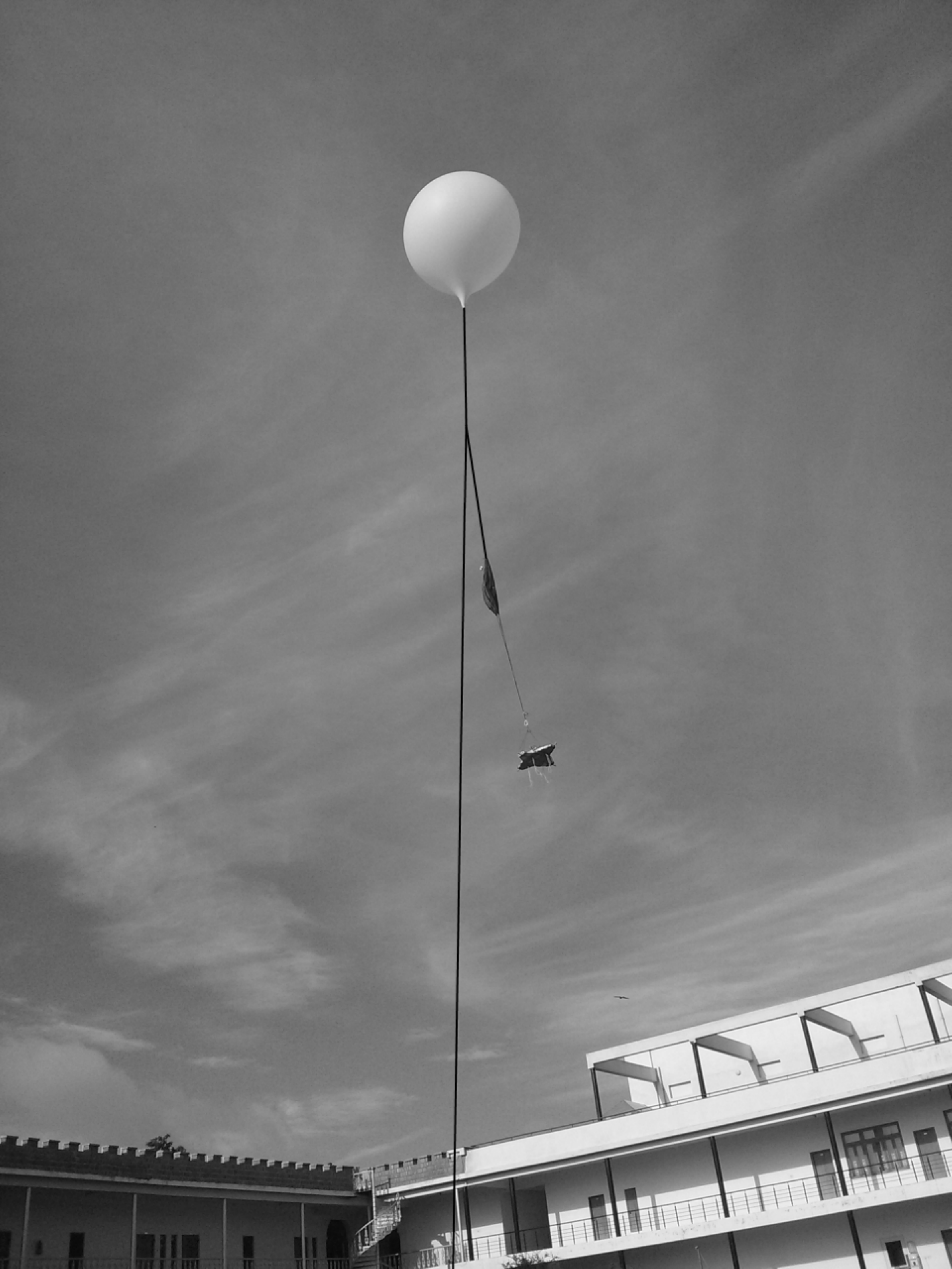}
\caption{The star-shaped platform is connected to the balloon through a parachute. The entire system is tethered to a spindle on the ground (not shown in figure) using a nylon tether of approximately 200 m length.}
\label{fig:tether}
\end{figure}

The payload, containing the pointing system with mounted attitude sensor, a digital camera, batteries and a radio tracker, was placed on a star-shaped platform. The platform 'wings' were tied to a parachute by the nylon rope, and the parachute was attached to balloon as shown in Fig.~\ref{fig:tether}, through a Nylon rope of approximately 7 m in length. In addition, the balloon was tethered through a nylon rope of approximately 200 m, which was fixed to a spindle on the ground. The aim of this launch was to point at and track the Sun, take images of the Sun and estimate pointing accuracy and stability. The UT starting time of the experiment and equatorial coordinates of the Sun were programmed into the ACS. The operation of the camera was adjusted to take the image every 5 seconds with an exposure time of 0.1 millisecond. This experiment was conducted three times over a period of four months. 

\listoffigures
\listoftables
\end{spacing}
\end{document}